\documentclass[a4paper,aps,twocolumn,prb,superscriptaddress,secnumarabic]{revtex4}
\usepackage{graphicx}
\usepackage{dcolumn}
\usepackage{color}
\usepackage{bm}
\usepackage[caption=false]{subfig}
\usepackage{hyperref}

\begin{document}
\newcommand{\mos}{MoS$_2$}
\preprint{APS/123-QED}

\title{Tuning band gaps in twisted bilayer MoS$_2$}

\author{Yipei Zhang}
\affiliation{Key Laboratory of Artificial Micro- and Nano-structures of Ministry of Education and School of Physics and Technology, Wuhan University, Wuhan 430072, China}
\author{Zhen Zhan}
\email{zhen.zhan@whu.edu.cn}
\affiliation{Key Laboratory of Artificial Micro- and Nano-structures of Ministry of Education and School of Physics and Technology, Wuhan University, Wuhan 430072, China}
\author{Francisco Guinea}
\affiliation{Fundaci\'on IMDEA Nanociencia, C/Faraday 9, Campus Cantoblanco, 28049 Madrid, Spain}
\affiliation{Donostia International Physics Center (DIPC)–UPV/EHU, E-20018 San Sebastián, Spain}
\author{Jose \'{A}ngel Silva-Guill\'{e}n}
\affiliation{Key Laboratory of Artificial Micro- and Nano-structures of Ministry of Education and School of Physics and Technology, Wuhan University, Wuhan 430072, China}
\author{Shengjun Yuan}
\email{s.yuan@whu.edu.cn}
\affiliation{Key Laboratory of Artificial Micro- and Nano-structures of Ministry of Education and School of Physics and Technology, Wuhan University, Wuhan 430072, China}


\begin{abstract}
In the emerging world of twisted bilayer structures, the possible configurations are limitless, which enables for a rich landscape of electronic properties.
In this paper, we focus on twisted bilayer transition metal dichalcogenides (TMDCs) and study its properties by means of an accurate tight-binding model.
We build structures with different angles and find that the so-called flatbands emerge when the twist angle is sufficiently small (around 7.3$^{\circ}$).
Interestingly, the band gap can be tuned up to a 2.2\% (51 meV) when the twist angle in the relaxed sample varies from 21.8$^{\circ}$ to 0.8$^{\circ}$.
Furthermore, when looking at local density of states we find that the band gap varies locally along the moir\`e pattern due to the change in the coupling between layers at different sites.
Finally, we also find that the system can suffer a transition from a semiconductor to a metal when a sufficiently strong electric field is applied. Our study can serve as a guide for the practical engineering of the TMDCs based optoelectronic devices.


\end{abstract}

\maketitle

\section{Introduction}
\label{sec:intro}
Although graphene has been known for some time now\cite{novoselov2004electric}, recently there has been a renewed interest in the properties of bilayer structures due to the discovery of strongly correlated effects in these structures at certain small (magic) twist angles.\cite{cao2018correlated}
This finding triggered a handful of experimental and theoretical studies in that kind of structures where not only strongly correlated effects such as superconductivity and quantum phase transitions\cite{bistritzer2011moire,laissardiere2010localization,laissardiere2012numerical,kim2017tunable,cao2018unconventional,yankowitz2019tuning,chen2019evidence,yoo2019atomic,kerelsky2019maximized,xie2019spectroscopic,jiang2019charge}, but also the existence of pseudo-magnetic fields due to the strain that the system can experience either by applying externally a mechanical strain \cite{guinea2010energy,yan2013strain} or due to intrinsic strain that appears in the moir\'e pattern because of the incommensurability of the superstructures.\cite{shi2020large}

Similarly to graphene, group V-B transition metal dichalcogenides (from now on TMDCs), are exfoliated materials that have an hexagonal structure and also change their electronic properties dramatically when lowering the number of layers to one.
Interestingly, in contrast to graphene, TMDCs are semiconducting and, moreover, the nature of this band gap depends on the number of layers changing from indirect to direct when the system goes from multi-layer to monolayer.\cite{roldan2017theory}
The fact that monolayer TMDCs present a direct band gap overcomes one of the major drawbacks of graphene for its integration to modern electronic and optoelectronic devices.
Furthermore, this band gap can be tuned by means of the so-called straintronics\cite{amorim2016novel} methods or electric fields.
Therefore, it seems like a natural and interesting step to study the electronic properties of twisted bilayer TMDCs.
Recently, a theoretical work using density functional theory (DFT) methods predicted the existence of flatbands in MoS$_2$ when the twist angle is sufficiently small.\cite{naik2018ultraflatbands} 
Furthermore, an experimental work was carried out on another twisted bilayer TMDCs, WSe$_2$, where they found such flatbands when achieving small twist angles.\cite{zhang2019flat}
Interestingly, it has been shown that the different environment surrounding the atoms due to the change in the stacking along the moir\`e pattern in heterobilayer TMDCs (a structure formed by a different TMDCs in each layer) entails a difference in the interlayer coupling, which results in a local change of the gap.\cite{zhang2017interlayer}
Nevertheless, a thorough study of the electronic properties and their possible tunability of the twisted bilayer TMDCs is still lacking.

In this work, we study the electronic properties of twisted bilayer MoS$_2$ and the possibility of tuning the band gap.
The paper is organized in the following way: 
We first show how to build the commensurate twisted bilayer TMDCs and the method used to compute their electronic properties. 
Then, we study the tunability of these properties by means of a change in the rotation angle, by the variation of the local interlayer couplings due to the different stackings in the moir\'e pattern or by applying an electric field to the system.

\section{THE COMMENSURATE BILAYERS}

\begin{figure}[t]
\centering
\includegraphics[scale=1]{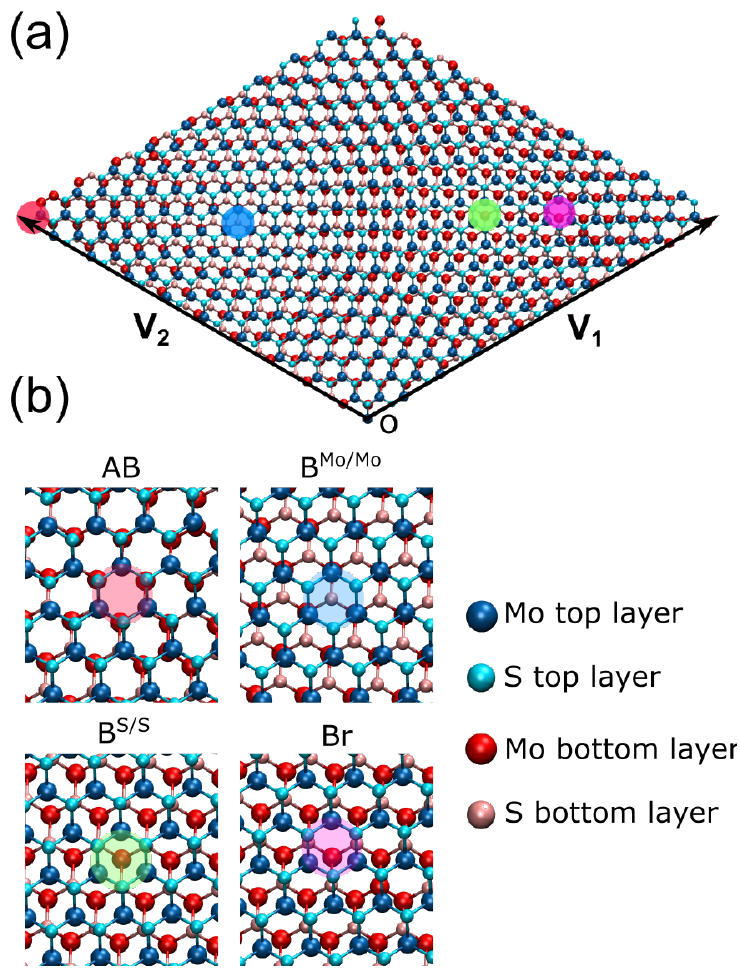}
\caption{(a) The moir\'{e} superlattice of twisted bilayer $ {\rm MoS_2}$ with rotation angle $ \theta = 3.5 ^{\circ}$. (b) Zoom in around the atomic structures of different high symmetry points. The  AB, $\rm B^{Mo/Mo}$, $\rm B^{S/S}$ and Br regions are highlighted by circles of different colors.}
\label{fig:atomic}
\end{figure}

We consider bilayer TMDCs, which are composed of two monolayers of MoS$_2$ rotated in the plane by an angle $\theta$. Since the two layers have the same lattice constant, following the same method as in twisted bilayer graphene, the moir\'{e} supercell can be constructed by identifying a common periodicity between the two layers \cite{de2012numerical}. 
We start from the 2H stacking ($\theta=0^\circ$) of \mos, that is with the Mo (S) atom in the top layer directly above the S (Mo) in the bottom layer, and choose the rotation origin, O, at an atom site. 
For top layer, we define a supercell with a basis vector $\mathbf{V_1}(n,m) = n\mathbf{a_1} + m\mathbf{a_2}$, being $\mathbf{a_1}$ and $\mathbf{a_2}$ the lattice vectors of single-layer ${\rm MoS_2}$, and $n$ and $m$ are non-negative integers with $n - m = 1$, which means that the supercell contains only one moir\'{e} pattern. 
For the bottom layer, a cell with the same size and rotated by an angle $\theta$ can be obtained with the basis vector $\mathbf{V_1'}(m,n)$. 
Then, the commensurate bilayers with the twist angle $\theta$ can be achieved by rotating the top cell with $\mathbf{V_1}$ by $\theta/2$ and rotating the bottom cell with $\mathbf{V_1'}$ by $-\theta/2$. 
The rotation angle is given by:
\begin{equation}
\cos\theta = \frac{n^2+4nm+m^2}{2(n^2+nm+m^2)}
\end{equation} 
The commensurate supercell contains $N = 6(n^2+nm+m^2)$ atoms, and the lattice vectors are $\mathbf{V_1}$ and $\mathbf{V_2} = -m\mathbf{a_1} + (n+m)\mathbf{a_2}$ with $|\mathbf{V_1}| = |\mathbf{V_2}| = \frac{a}{2\sin(\theta/2)}$, where $a=3.16$~\AA~is the lattice constant of the single-layer $\mathrm{ MoS_2}$ \cite{roldan2014electronic}. 
Fig. \ref{fig:atomic}(a) shows a twisted bilayer \mos~ structure with the twist angle $\theta = 3.5^{\circ}$, which is obtained with $n = 10$ and $m = 9$. 
The moir\'{e} superlattice contains 1626 atoms. 
In a supercell with relatively small twist angle there are several high-symmetry stacking patterns, for instance, AB, $\mathrm{ B^{Mo/Mo}}$ and $\mathrm{ B^{S/S}}$.
In the AB stacking, the Mo atoms of layer 1 are over the S atoms of layer 2 and the S atoms of layer 1 are over the Mo atoms of layer 2. For $\rm B^{Mo/Mo}$, Mo of layer 1 are over Mo of layer 2 and  all S of one layer are located in the center of hexagons of the other layer. For $\rm B^{S/S}$, S of layer 1 are over S of layer 2 and all Mo of one layer are in the center of hexagons of the other layer. 
The Br site is located at one third of the $\rm B^{S/S}$ -- AB path.  
All of these special sites are illustrated in Fig. \ref{fig:atomic}(b).

\section{NUMERICAL METHOD}

Ultraflatbands at the valence band edge were discovered theoretically in twisted bilayer ${\rm MoS_2}$ \cite{naik2018ultraflatbands,Naik2020,PhysRevB.102.081103}. 
Up to now, the largest system of this kind calculation using first-principles methods contains 4902 atoms, which corresponds to a twist angle of $2.0 ^{\circ}$.
Although it is possible to perform calculations on larger systems, there are some limitations due to the computational resources when the twist angle becomes smaller since the number of atoms increases sharply. 
A systematic study of these larger systems can be more easily done by utilizing a tight-binding method. 
For instance, the system with the electronic properties calculated by diagonalization in reciprocal space contains up to 59644 orbitals, which corresponds to $\theta = 1^{\circ}$. 

In this paper, we will use another approach, the tight-binding propagation method (TBPM), to investigate the electronic properties of the twisted bilayer $\rm MoS_2$. \textcolor{red}{The TBPM is based on the numerical solution of the time-dependent Schr\"{o}dinger equation without any diagonalization \cite{yuan2010modeling}. Both memory and CPU costs scale linearly with the system size. Therefore, the TBPM can tackle systems with the number of orbitals as large as ten million, for instance, extremely tiny twist angles in twisted bilayer graphene \cite{shi2020large} and bilayer graphene quasicrystals \cite{yu2019dodecagonal}. 
More importantly, defects, magnetic and electric fields can be easily implemented in this approach. 
We briefly outline the main formalism of using the TBPM to calculate the density of the states. TBPM starts with a random superposition of basis function $|\phi_0 \rangle = \sum c_i |a_i \rangle $, where {$c_i$} are random complex numbers and {$|a_i\rangle$} are basis states of the calculated sample. Then, by solving the time-dependent Schr\"{o}dinger equation, the DOS is obtained from the Fourier transform of the time-dependent correlation function:
$d(\epsilon) = \frac{1}{2\pi}\int^{+\infty}_{-\infty}e^{i\epsilon \tau} \langle \phi_0|e^{-iH\tau/\hbar} |\phi_0\rangle d\tau$, where $H$ is the Hamiltonian of the system. In this method, the accuracy is determined by the number of orbitals in the sample and can be increased by using larger samples or averaging results from different random initial states. The number of the time integration steps determines the energy resolution. The larger the system, the more accurate the calculated results. Such method has been implemented in our home-made program Tipsi (Tight-binding propagation simulator) where density of the states, local density of states, quasieigenstates and many other electronic and optical properties can be easily obtained once the Hamiltonian of the system is given. 
}

\begin{figure*}[t!]
\centering
\includegraphics[width=0.9\textwidth]{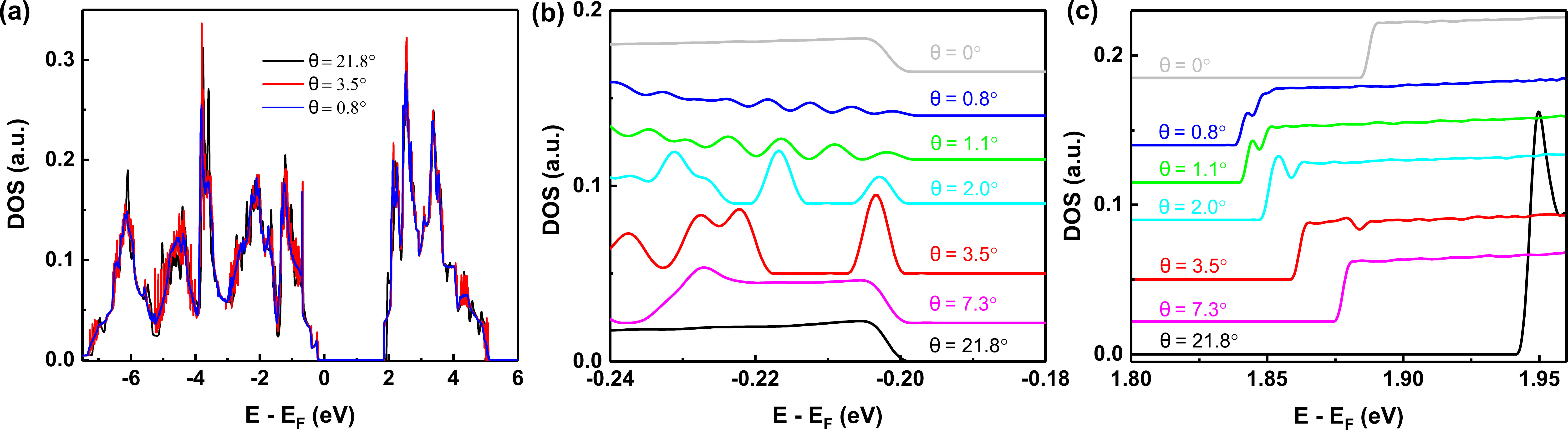}
\caption{(a) The calculated DOS of rigidly twisted bilayer $ {\rm MoS_2}$ with different twist angles. (b) and (c) The detailed changes of DOS near the valence  and conduction band extrema, respectively.}
\label{fig:DOS}
\end{figure*} 

In order to calculate the electronic band structures of twisted bilayer TMDCs, we use an accurate tight-binding Hamiltonian introduced in Ref. \onlinecite{fang2015ab}. 
The minimum atomic orbital basis to correctly describe monolayer TMDCs are the five \textit{d} orbitals of the transition-metal atom and three \textit{p} orbitals of each of the two chalcogen atoms.\cite{cappelluti2013tight,silva2016electronic} 
This model well reproduces the band structure calculated using  DFT with \emph{GW} quasi-particle correction in the low energy region. 
For bilayer TMDCs, the total Hamiltonian can be written as\cite{roldan2014momentum,fang2015ab}:
\begin{equation}
\hat H = \hat H_1^{(1L)}+\hat H_2^{(1L)}+\hat H_{int}^{(2L)},
\label{hal}
\end{equation}
where $ \hat H_{1(2)}^{(1L)}$ is the eleven-orbital single layer Hamiltonian and $\hat H_{int}^{(2L)}$ is the interlayer Hamiltonian.
$\hat H_{1(2)}^{(1L)}$ contains the on-site energy, the hopping terms between orbitals of the same type at first-neighbor positions and the hopping terms between orbitals of different type at first- and second-neighbor positions. 
The  interlayer hopping Hamiltonian only includes the interaction between the chalcogen atoms at the interface of the bilayer:
\begin{eqnarray}
\hat H_{int}^{2L} &=& \displaystyle\sum_{p_i',\mathbf r_2,p_j,\mathbf r_1}\hat \phi_{2,p_i'}^\dagger(\mathbf r_2)t_{p_i',p_j}^{(LL)}(\mathbf r_2-\mathbf r_1)\hat \phi_{1,p_j}(\mathbf r_1), \nonumber\\ 
&& + h. c. 
\end{eqnarray}
where $\hat \phi_{i,p_j}$ is the $p_j$ orbital basis of $i$-th monolayer. 
Within the Slater-Koster approximation, the hopping terms can be expressed as:
\begin{equation}
t_{p_i',p_j}^{(LL)}(\mathbf r) = (V_{pp,\sigma}(r)-V_{pp,\pi}(r))\frac{r_ir_j}{r^2}+V_{pp,\pi}(r)\delta_{i,j},
\end{equation}
where $r=|\mathbf r|$ and the distance-dependent Slater-Koster parameter is:
\begin{equation}
V_{pp,b}=\nu_be^{[-(r/R_b)^{\eta_b}]},
\end{equation}
where $b=\sigma,\pi$, $\nu_b$, $R_b$ and $\eta_b$ are constant values that depend on the specific of the chalcogen interlayer interactions, of which values are taken from the Ref. \onlinecite{fang2015ab}. 
In our calculations, we only include the interlayer hopping terms between a pair of chalcogen atoms that are separated by a distance smaller than 8 \AA~.

\textcolor{red}{In all the calculations, we use a large enough system with more than 10 million orbitals to ensure the convergence of the results. For instance, the number of orbitals in the unit cell of twisted bilayer $\rm MoS_2$ with $\theta=2.0^\circ$ is 17974. To perform the calculation with the TBPM, we use a large sample containing $31\times31$ unit cells. The time steps are set to 4096, which gives an energy resolution of 1.8 meV.  Periodic boundary conditions are used in the simulation. Furthermore, we can use TBPM to obtain the map of the amplitudes of the quasieigenstates which has been shown to be in agreement with the measured dI/dV mapping in experiments (for instance, the results in Ref. \onlinecite{shi2020large}).
Note that the band structure calculations in Sec. \ref{sec:elecfield} are performed by standard diagonalization of the Hamiltonian in Eq. (\ref{hal}).}

\section{Results and Discussion}

\subsection{Tuning the band gap by rotation angle}

It has been proven that the twist angle has a significant influence on the electronic properties of twisted bilayer TMDCs \cite{huang2014probing,puretzky2016twisted,van2014tailoring,liu2014evolution,yeh2016direct}. 
All these studies are mainly focused on large rotation angles. 
Interestingly, ultraflatbands have been detected in low-angle twisted bilayer ${\rm MoS_2}$. These flatbands provide good platform for exploring new physical phenomena, for instance, the Mott-insulating phase at half-filling of the band \cite{naik2018ultraflatbands,cao2018correlated}. 
This leaves important questions unaddressed: Are there ultraflatbands in twisted TMDCs with tiny twist angle? 
What exotic features will be found in low-angle twisted bilayer TMDCs?

\begin{figure}[t!]
\centering
\includegraphics[width=0.43\textwidth]{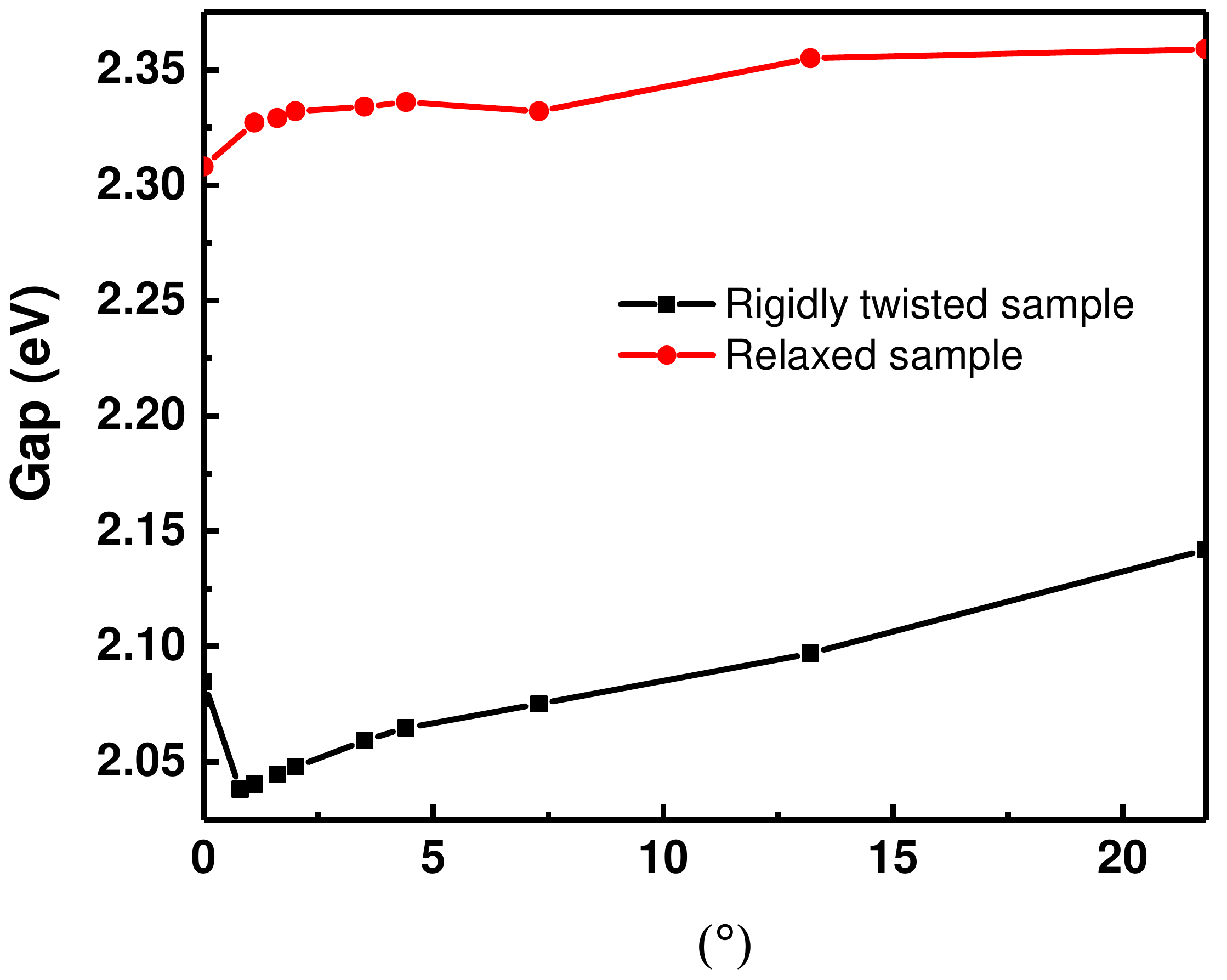}
\caption{The band gap $E_g$ between the CBM and VBM of twisted bilayer $ {\rm MoS_2}$ with different rotation angles. The black and red lines are for unrelaxed and relaxed cases, respectively.}
\label{gap1}
\end{figure}

In this part, we study the density of states of twisted bilayer MoS$_2$ with various rotation angles. The smallest rotation angle that we calculate is $0.8^{\circ}$ which results in a moir\'{e} pattern that contains 29526 atoms (108262 orbitals). It is far beyond the ability of state-of-the-art first-principles methods and tight-binding methods where the electronic structure is calculated by using diagonalization methods. 
The DOS of rigidly twisted bilayer ${\rm MoS_2}$ with angles changing from $0^{\circ}$ to $21.8^{\circ}$ are plotted in Fig.\ref{fig:DOS} (a). 
It is clear that the DOS varies significantly depending on the angle, especially for the DOS deep into the valence band, which is in good agreement with the calculated results in Ref. \onlinecite{carr2017twistronics}. 
More interesting things happen near the band edges. 
In order to investigate this, the detailed evolution of DOS near the band edge is illustrated in Fig.\ref{fig:DOS}~(b) and (c). 
{\color{red}We see clearly that, except for the $0^\circ$ that corresponds to the 2H stacking, as the twist angle decreases, the conduction band edge energy decreases, and the energy gap decreases.} That is, the band gap can be engineered through the control of the rotation angle. As shown in Fig. \ref{gap1} (the black line), the band gap reduces by 104 meV (around 5\% change) when changing the rotation angle from $21.8^{\circ}$ to $0.8^{\circ}$.  
Note that, for samples with small twist angle, some energy peaks appearing near the valence band edges correspond to the detected ultraflatbands. For instance, in the DOS of the twisted bilayer ${\rm MoS_2}$ with twist angle $\theta = 3.5^\circ$, the peak located at -0.2 eV corresponds to the ultraflatband discovered in Refs. \onlinecite{naik2018ultraflatbands,Zhen2020flat}.
\textcolor{red}{In principle, for rigidly twisted $\rm MoS_2$ with rotation angles below a crossover value $\theta^\ast \approx 7^\circ$, the isolated flat band emerges, and the states of the flatband in the VBM are localized in the $\rm B^{S/S}$ region (see Sec. \ref{sec:localization}).}
This is consistent with reported experimental and DFT results\cite{naik2018ultraflatbands,zhang2019flat}.

\textcolor{red}{Next, we investigate the influence of the lattice relaxation on the band gap of twisted bilayer $\rm MoS_2$ with various rotation angles. The structural relaxations are performed with the LAMMPS\cite{plimpton1995fast,lammps} package where the Stillinger-Weber potential\cite{jiang2015parametrization} for the interactions between atoms within the layer and the Lennard-Jones potential\cite{rappe1992uff} for the interlayer interactions are implemented. The minimizations are performed using the conjugate gradient method with the energy tolerance being $1e^{-15}$ eV. The relaxed sample is assumed to keep the period of the rigidly twisted $\rm MoS_2$. 
It has been shown that the relaxation results in a maximum of a 1\% displacement in the in-plane directions\cite{li2020imaging}. For simplicity, in the relaxed simulations, we neglect the effect of the relaxation on the intralayer hopping. Such simulation gives a qualitatively study of the modulation of the band gap by the interlayer hopping. As we can see from the red line in the Fig. \ref{gap1}, the band gap still reduces with the rotation angle. It declines 51 meV from $\theta=21.8^\circ$ to $0^\circ$. Different from the rigid case, the 2H stacking has the minimum band gap. Moreover, the lattice relaxation increases the band gap for all the twisted samples. In the tiny twist angle, the reduction of the band gap is compensated by the lattice relaxation. As we know the relaxation effect is more significant in small rotation angle samples. So we can see from the red line in Fig. \ref{gap1}, the band gap changes more smoothly in the small twist angle.  
Such relaxation effect can be suppressed when placing the sample on a hBN substrate\cite{dean2010boron,li2020imaging}.}

\subsection{Tuning the band gap by interlayer coupling at different high-symmetry stacking points}\label{sec:localization}

\begin{figure}[h!]
\centering
\includegraphics[width=0.43\textwidth]{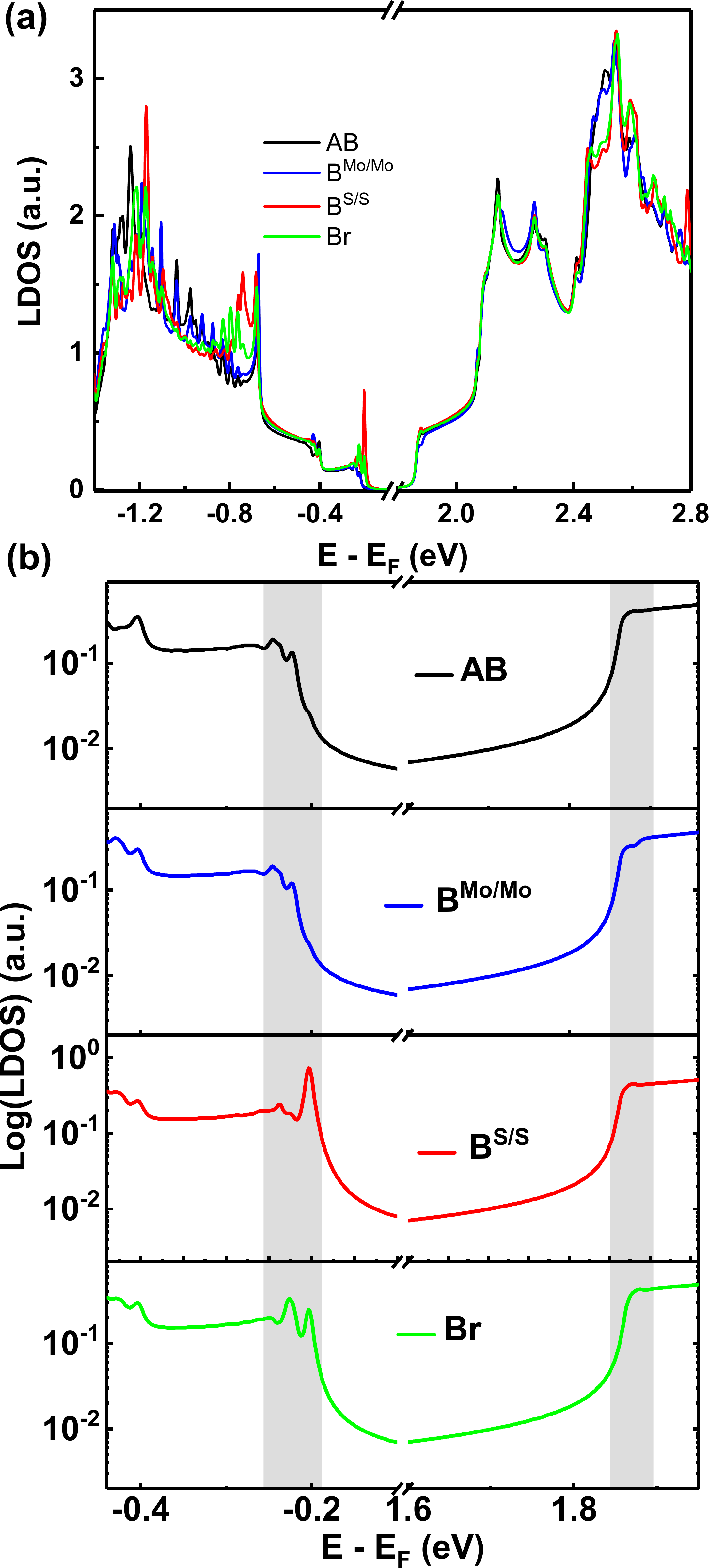}
\caption{(a) The calculated LDOS at the AB, $\rm B^{Mo/Mo}$, $\rm B^{S/S}$ and Br regions in rigidly twisted bilayer $ {\rm MoS_2}$ sample with $ \theta = 3.5 ^{\circ}$. (b) The logarithmic scales of the LDOS near the valence and conduction band edges.}
\label{fig:ldos}
\end{figure}

\begin{figure}[t!]
\centering
\includegraphics[width=0.45\textwidth]{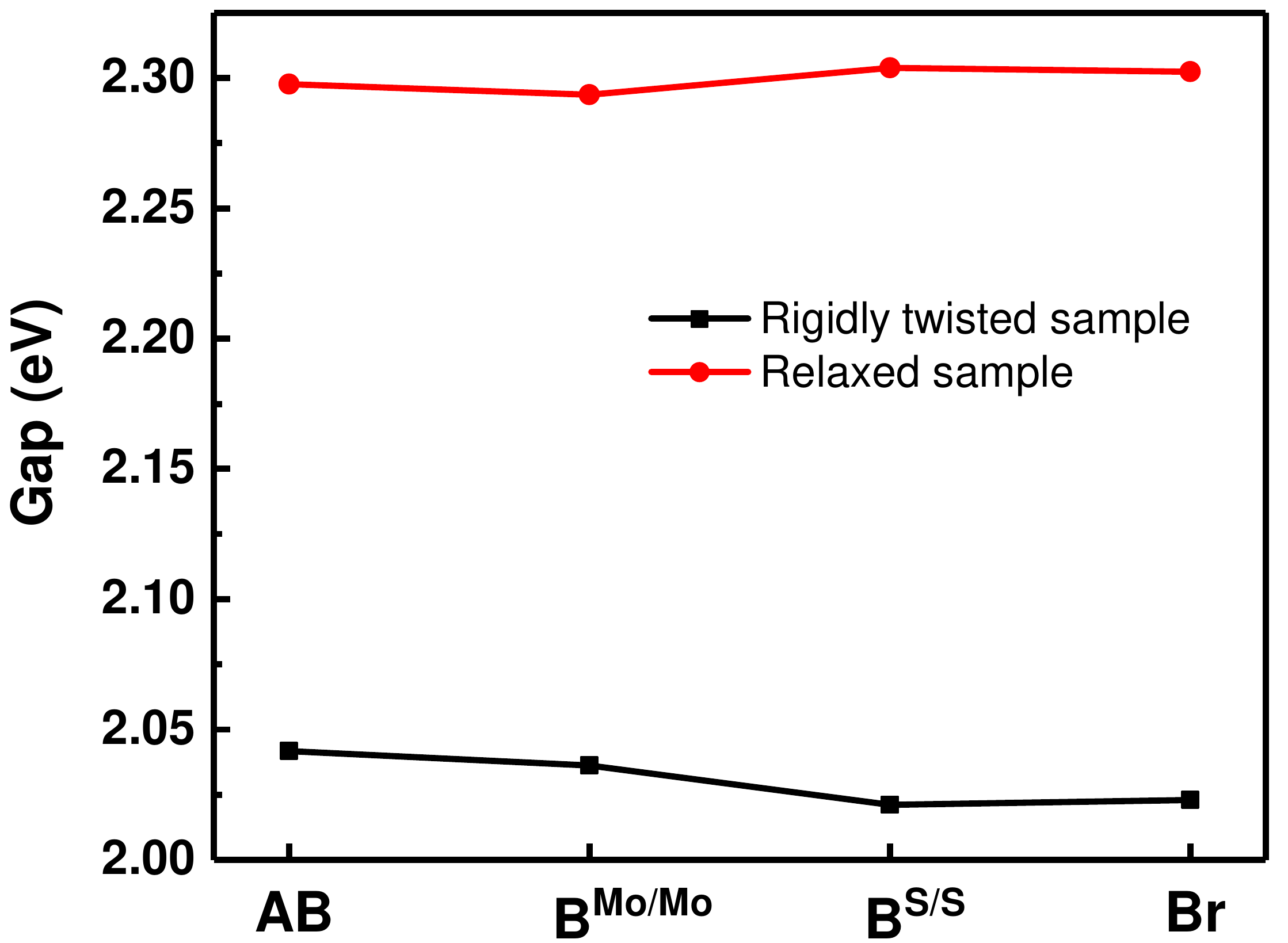}
\caption{The local band gap $E_g$ of the AB, $\rm B^{Mo/Mo}$, $\rm B^{S/S}$ and Br regions in twisted bilayer $ {\rm MoS_2}$ with $ \theta = 3.5 ^{\circ}$. The black and red lines are for the rigid and relaxed cases, respectively.}
\label{fig:gap}
\end{figure}

It has been shown experimentally that, for heterostructures composed of monolayers of two different TMDCs, the local band gap is periodically modulated by the interlayer coupling at different high-symmetry points with an amplitude of $\sim 0.15$ eV \cite{zhang2017interlayer}. 
Can the interlayer coupling be used as a parameter to tune the local band gap for the twisted homobilayer TMDCs (the heterostructure composed of the same monolayers TMDCs)? 
To answer this question, we calculate the local density of states for the twisted bilayer ${\rm MoS_2}$ with $ \theta = 3.5 ^{\circ}$ at the high symmetry stacking points AB, ${\rm B^{Mo/Mo}}$, ${\rm B^{S/S}}$ and Br (illustrated in Fig.\ref{fig:atomic}). 

The results are shown in Fig.\ref{fig:ldos}(a). 
Similar to the DOS in Fig. \ref{fig:DOS}, the interlayer coupling changes also significantly the LDOS deep into the valence band. 
The details of the LDOS near the valence band maximum (VBM) and conduction band minimum (CBM) are plotted in Fig. \ref{fig:ldos}(b). 
In the conduction band, the LDOS near the band edge are similar for the four high-symmetry sites. 
On the contrary, the VBM has a remarkable change due to the different interlayer coupling at the four points, which can be seen more clearly in the logarithm of the LDOS illustrated in Fig. \ref{fig:ldos}(b). 
Furthermore, we find in Fig.\ref{fig:ldos} that the ultraflatband signature, which corresponds to the peak with energy $\sim$ -0.2 eV, only appears at ${\rm B^{S/S}}$ and Br points. 
We can also see that the sharpest peak appears in ${\rm B^{S/S}}$ sites. 
This is expected since at this position the top layer S atom sits directly above a bottom layer S atom, which gives the strongest interlayer interaction, given the fact that in our tight-binding model, interlayer coupling originates from hopping between S atoms in different layers. 
The absence of signals of flatband on other areas indicates that the localization of the electronic states of the flatband is around the ${\rm B^{S/S}}$ site, which is in accordance with the localization of the VBM wave function in the rigidly twisted sample in Ref. \onlinecite{naik2018ultraflatbands}. 
The local energy gap at different stacking regions is shown in the black line of Fig. \ref{fig:gap}.
We can see how the band gap changes locally depending on the specific site.
At ${\rm B^{S/S}}$, which has the strongest interlayer coupling, we find the minimum local energy gap. 
The local band gap is modulated periodically with an amplitude of $\sim 35$ meV. 
The evolution of such site-dependent local band gap is in agreement with the experimental results reported in heterostructure TMDCs \cite{zhang2017interlayer}. \textcolor{red}{In the relaxed case, as shown in the red line of Fig. \ref{fig:gap}, the minimum band gap is located at the AB and $\rm B^{Mo/Mo}$ regions, and the difference of the band gap at different high-symmetry points is reduced to 5 meV.} 
All in all, an important consequence of the interlayer coupling in the moir\'{e} supercell is the tuning the local band gap at different stacking points.

\begin{figure}[t!]
\centering
\includegraphics[width=0.45\textwidth]{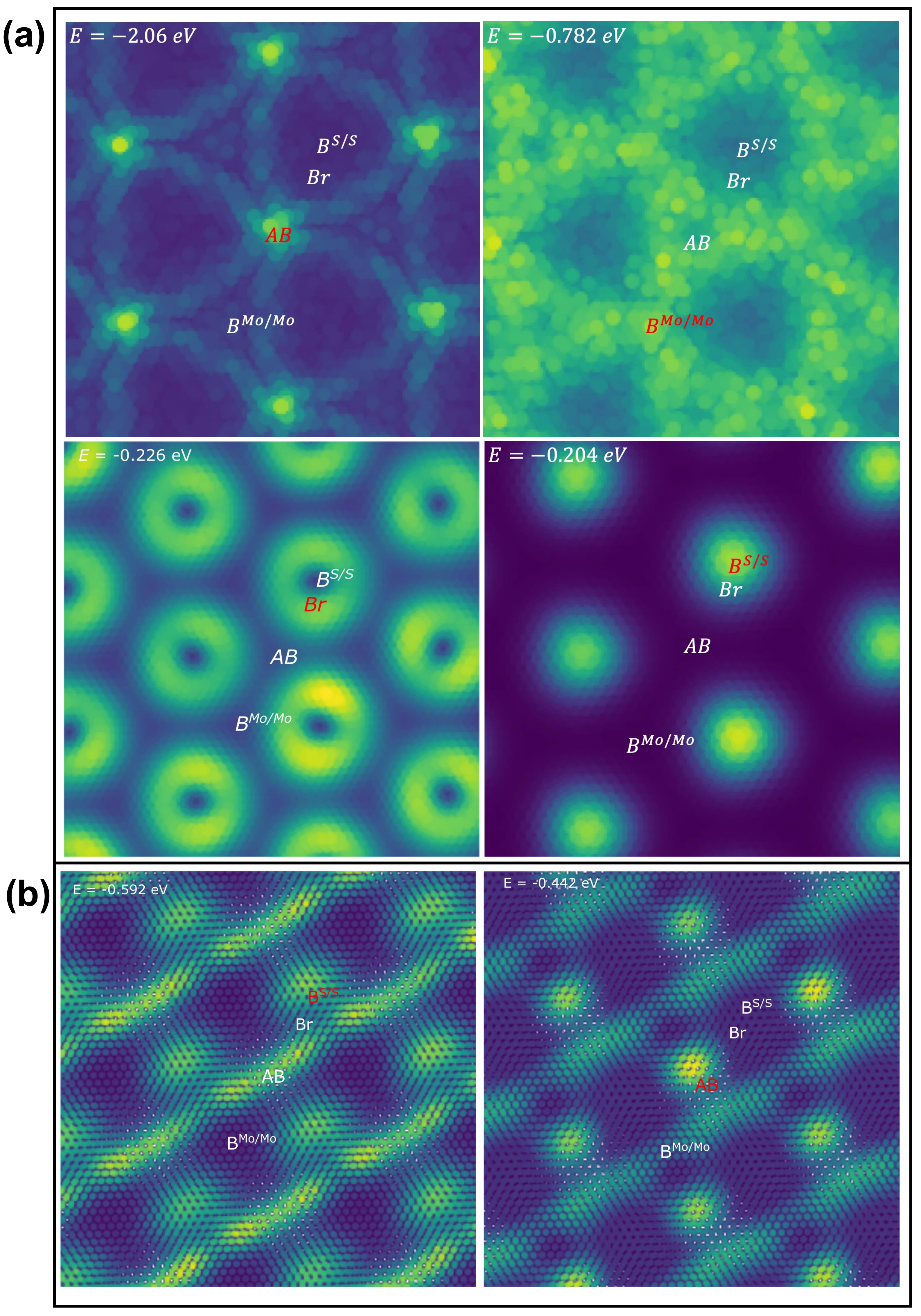}
\caption{Calculated LDOS mapping with different energies of the twisted bilayer $ {\rm MoS_2}$ with $ \theta = 3.5 ^{\circ}$ for the (a) rigid and (b) relaxed cases. The corresponding energies and the four special sites AB, $\rm B^{Mo/Mo}$, $\rm B^{S/S}$ and Br are labeled in each image.}
\label{fig:mapping}
\end{figure}
The periodic variation of the local electronic structure as a consequence of different interlayer couplings is also visualized more clearly looking at the energy dependence of the spatial distribution of LDOS plotted in Fig. \ref{fig:mapping}.  
In the rigid cases, for instance, at a high negative energy of -2.06 eV, where a peak appears in the LDOS of the AA site, the spectra at that same position is higher than that of the other three high-symmetry points. 
However, at energy around -0.2 eV, where the spectral feature of the AA site is out of the tunneling range, the intensity of the AA site changes from a bright feature to a deep hole, whereas the ${\rm B^{S/S}}$ and Br sites have the highest spectral at energies -0.226 eV and -0.204 eV, respectively.
As we discuss previously, all the states are localized around the ${\rm B^{S/S}}$ site at -0.2 eV. The continuous evolution of the local electronic spectral by different energies at different sites  also occurs for positive energies (not shown here).  \textcolor{red}{On the other hand, for the relaxed sample, as shown in the Fig. \ref{fig:mapping}(b), the first flatband (-0.442 eV) in the VBM is mainly located at the AB and $\rm B^{Mo/Mo}$ sites. The localization of the state at the $\rm B^{Mo/Mo}$ and the breaking of the $C_3$ symmetry are due to the fact that we neglect the effect of in-plane movements on the intralayer hopping. For higher energies, such as -0.592 eV, some states are localized at the $\rm B^{S/S}$ site.} All in all, all these results show a periodic charge density modulation at different energies over large areas in both rigid and relaxed cases, which can be detected experimentally using scanning tunneling microscopy dI/dV mapping.

\subsection{Tuning the band gap by applying an electric field}\label{sec:elecfield}

\begin{figure}[ht]
\centering
\includegraphics[width=0.45\textwidth]{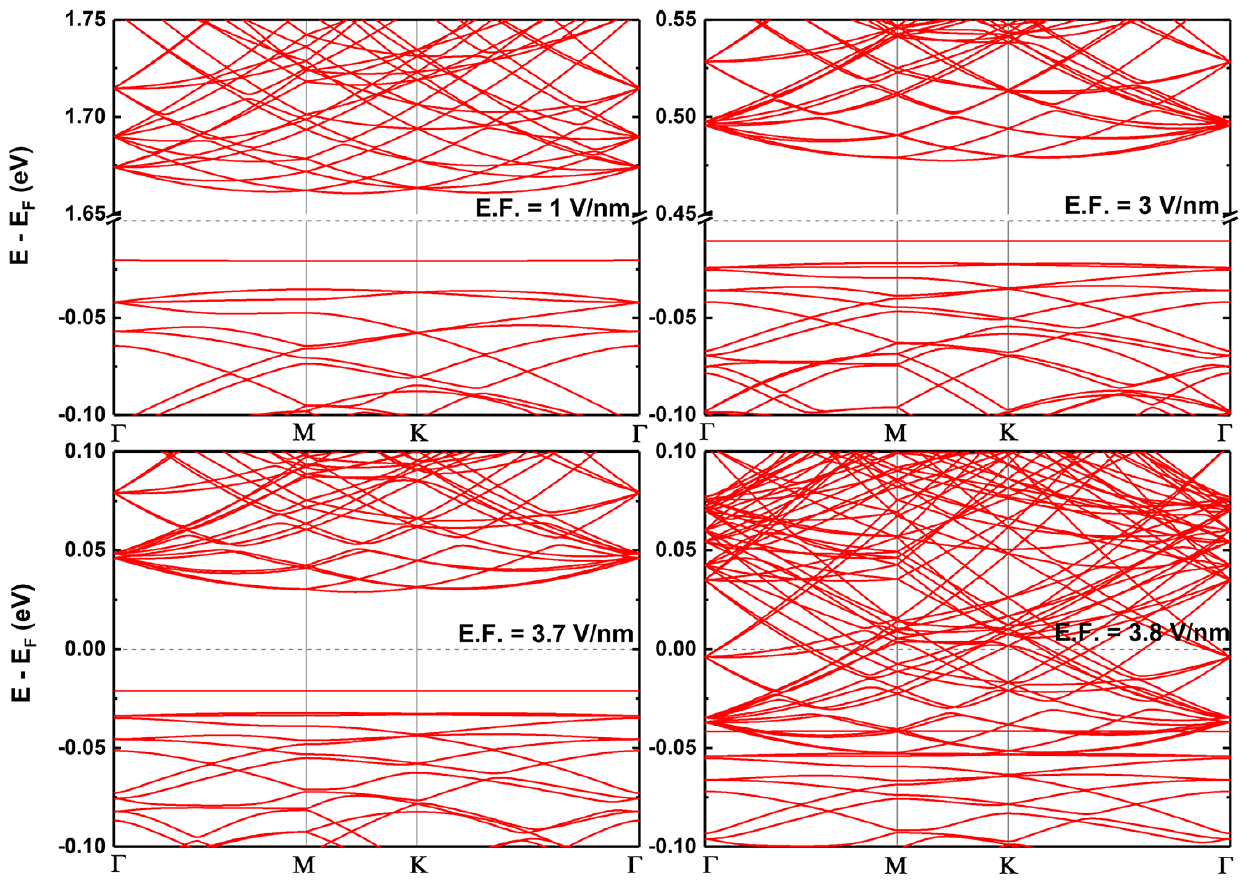}
\caption{Band structure of rigidly twisted bilayer ${\rm MoS_2}$ with $ \theta = 3.5 ^{\circ}$ along $\Gamma$ - K - M - $\Gamma$ direction in reciprocal space as a function of the applied external electric field. \textcolor{red}{The band structure are obtained by exact diagonalization of the Hamiltonian in Eq. (\ref{hal}). Note that the vertical axis scales are different in each panel.}}
\label{fig:BS}
\end{figure}

As it has been shown, a vertical electric field can open a bandgap in bilayer graphene \cite{zhang2009direct,mak2009observation}.
Furthermore, bilayer TMDCs can suffer a transition from semiconductor to metal when the applied field is strong enough  \cite{ramasubramaniam2011tunable,liu2012tuning,zhang2019interface}. 
However, up to the date the effect of the electric field and the possible modulation of the band gap in small angle twisted TMDCs has not been studied.
Since all of these materials could be integrated into new electronic devices where a gate is applied, the study of this effect is of much interest.
In this part, we investigate the band gap tuning in twisted bilayer TMDCs by an external electric field applied perpendicularly to the layers, in particular, the twisted bilayer ${\rm MoS_2}$ with $ \theta = 3.5 ^{\circ}$.

\begin{figure}[t]
\centering
\includegraphics[width=0.45\textwidth]{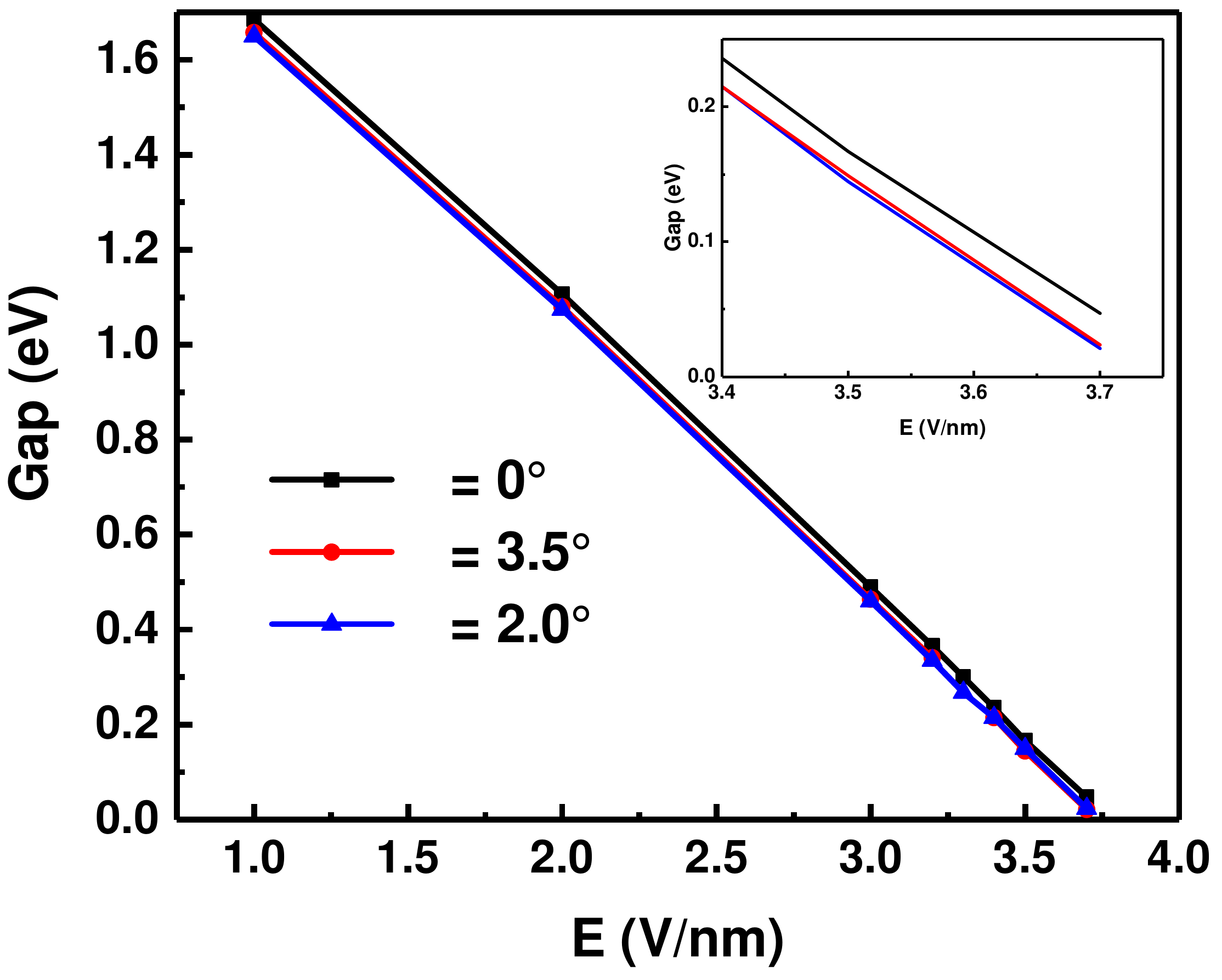}
\caption{The band gap of rigidly twisted bilayer ${\rm MoS_2}$ with different rotation angles as a function of the applied external electric field.}
\label{fig:gap_bs}
\end{figure}

Fig. \ref{fig:BS} shows the band structure of rigidly twisted bilayer ${\rm MoS_2}$ with $ \theta = 3.5 ^{\circ}$ under four different strengths of external electric fields perpendicular to the layers. 
The band gap is driven linearly to zero with electric field $E$ increases and the system changes from semiconductor to metal when $E$ is large enough.
This can be easily understood thanks to the so-called giant Stark effect (GSE)\cite{khoo2004tuning}. Due to the redistribution of the charge density on different layers when an electric field is applied, bands belonging to different layers are separated from each other, which results in the reduction of the energy gap. 
This same effect is also found in 2H stacking bilayer TMDCs\cite{ramasubramaniam2011tunable,liu2012tuning} and large angle twisted bilayer WS$_2$\cite{zhang2019interface}. 
The evolution of the band gap as a function of $E$ for twisted bilayer ${\rm MoS_2}$ with three different rotation angles is plotted in Fig. \ref{fig:gap_bs}. Since the difference of the band gap in the three twist angles are quite small, the threshold values where the system becomes metallic do not change significantly with the twist angle. \textcolor{red}{The band gaps in the relaxed samples are larger than that of the same rigid ones. 
Consequently, in the relaxed system, we would need a higher electric field to close the band gap.}

\section{Conclusion}

In this paper we have studied the electronic properties of twisted bilayer \mos~ and their possible tunability by means of an accurate TBPM.
We have seen that the flatband appears when reaching angles sufficiently small. 
Interestingly, we can tune the value of the gap up to a 2.2\% just by changing the rotation angle in relaxed systems.
Furthermore, the gap is modulated at different high-symmetry positions of the structure due to the different interlayer couplings that appear.
We have also shown that another effective method to tune the band gap is by applying a perpendicular electric field,
In fact, the band gap diminishes with increasing electric field and the system can undergo a transition from semiconductor to metal when the field is high enough.

\section*{Acknowledgement}
\addcontentsline{toc}{section}{Acknowledgement}
This work was supported by the National Science Foundation of China under Grant No. 11774269. F. G. acknowledges support by funding from the European Commision, under the Graphene Flagship, Core 3, grant number 881603, and by the grants NMAT2D (Comunidad de Madrid, Spain),  SprQuMat and SEV-2016-0686, (Ministerio de Ciencia e Innovación, Spain). 
Numerical calculations presented in this paper have been performed on the supercomputing system in the Supercomputing Center of Wuhan University.

\bibliography{bandgap.bib}

\end{document}